\documentclass{webofc}
\usepackage[varg]{txfonts}   % Web of Conferences font
\usepackage{hyperref}
\begin{document}
\title{Generating Galaxy Clusters Mass Density Maps from Mock Multiview Images via Deep Learning}

\author{\lastname{Daniel de Andres}\inst{1}\fnsep\thanks{email:daniel.deandres@uam.es} \and
        \lastname{Weiguang Cui}\inst{1,2} \and
        \lastname{Gustavo Yepes}\inst{1} \and
        \lastname{Marco De Petris}\inst{3} \and 
        \lastname{Gianmarco Aversano}\inst{4} \and
        \lastname{Antonio Ferragamo}\inst{3,5} \and 
        \lastname{Federico De Luca}\inst{3} \and 
        \lastname{A. Jiménez Muñoz}\inst{1} 
        %\lastname{Friends of Friends}\inst{1}
        % etc.
}

\institute{Departamento de Física Teórica and CIAFF, Modulo 8 Universidad Autónoma de Madrid, 28049 Madrid, Spain.
\and
           Institute for Astronomy, University of Edinburgh, Blackford Hill, Edinburgh, EH9 3HJ, UK
\and
           Dipartimento di Fisica, Sapienza Universitá di Roma, Piazzale Aldo Moro, 5-00185 Roma, Italy
\and 
           EURANOVA, Mont-Saint-Guibert, Belgium
\and 
           Instituto de Astrofísica de Canarias (IAC) La Laguna, 38205, Spain
          }

\abstract{%
Galaxy clusters are composed of dark matter, gas and stars. Their dark matter component, which amounts to around 80\% of the total mass, cannot be directly observed but traced by the distribution of diffused gas and galaxy members. In this work, we aim to infer the cluster's projected total mass distribution from mock observational data, i.e. stars, Sunyaev-Zeldovich, and X-ray, by training deep learning models. To this end, we have created a multiview images dataset from {\sc{The Three Hundred}} simulation that is optimal for training Machine Learning models. We further study deep learning architectures based on the U-Net to account for single-input and multi-input models. We show that the predicted mass distribution agrees well with the true one.
}
\maketitle
\section{Introduction}
\label{intro}

Clusters of galaxies are laboratories for studying both cosmology and astrophysics since they are the biggest gravitationally-bound objects in The Universe  \cite{Kravtsov:2012}. Particularly, cosmological parameters can be estimated by studying the number of galaxy clusters and their evolution in mass and redshift (e.g. \cite{Salvati:2022} ), and therefore, accurately estimating their total mass is of paramount importance. The main components of a cluster of galaxies are Dark Matter (DM), which amounts to around 80\% of its mass, stars, mainly confined in galaxies, and hot gas in the Intra-Cluster Medium (ICM). 

The ICM is observed from the electrons' Bremsstrahlung effect in  X-rays by,  e.g.,  eROSITA \citep{eROSITA} or by  XMM-Newton within the CHEX-MATE \citep{CHEX-MATE} project. In addition, the ICM is also targeted at microwave frequencies through the Sunyaev-Zel'dovich (SZ) effect due to inverse Compton scattering of photons from the  cosmic microwave background. Different experiments such as the Planck satellite \citep{Planck:PSZ2} and the South Pole Telescope (SPT; \citep{SPT}) are providing insight into the physical properties of the intracluster gas.

On the other hand, large  galaxy surveys from optical telescopes are providing important information  on the   cluster galaxy members, such as the Sloan Digital Sky Survey (SSDS; \citep{SSDS}). However, the DM component, which amounts to  80 per cent of mass, cannot be directly observed. Therefore, the total mass of a galaxy cluster is usually inferred from ICM mass proxies or from its galaxy members' dynamics. Only by performing Gravitational Lensing observations, one can estimate, by applying theoretical models, its mass density. These lensing observations are very scarce, only around one hundred galaxy clusters are observed \cite{Herbonnet:2020} in comparison to a few thousand that are currently available for ICM surveys.

Nevertheless, cosmological simulations show that all these mass proxies are biased, due to the  theoretical assumptions made, such as the hypothesis  that the intracluster gas is in  Hydrostatic Equilibrium (HE), see for example \cite{Gianfagna:2021,Gianfagna:2023}. To address this problem, novel Machine Learning models (ML) have recently been used to estimate masses of galaxy clusters \cite{Ntampaka:2019,Ho:2019,Gupta:2020,Kodi:2020,Yan:2020,Kodi:2021,Gupta:2021,Ho:2021,Ho:2023,Krippendorf:2023}, which are free from theoretical assumptions and yield "bias-free" results. All of these Machine Learning models are based on Deep Learning (for a review see, e.g. \cite{Huertas:2023}), which makes use of convolutional layers in the neural network architecture to extract information from multi-dimensional data such as images from astronomical instruments and it utilises that information to estimate the mass. We have to point out that all these works are using supervised training that rely on big datasets with known total mass of the galaxy clusters in order to train Deep Learning models.  Therefore, one has to resort to the use of cosmological simulations, where a rich variety of galaxy clusters with different dynamical states \citep{DeLuca:2021} are available, and the underlying true total mass distribution is completely known. Nevertheless, training from simulations does not necessarily imply that the trained model can be directly used for estimating the masses of galaxy clusters in real observations from both ICM tracers or cluster galaxy members.   Only recently, this step -- directly estimating the galaxy cluster masses from observation --  was first established by de Andres et al. \cite{deAndres:2022}. However, all these previous studies focused on the estimation of the mass inside a certain overdensity (radius) referred to as $R_{200}$ or $R_{500}$. In Ferragamo et al \cite{Ferragamo:2023}, we  made another step forward -- predicting the total density profile based on the estimated masses at  different radii. With more information estimated for the cluster mass distribution, we can acquire more knowledge of the cluster's internal properties, thus their formation and evolution. For example, the concentration of cluster is directly inferred by the ML model from the SZ cluster images  \citep{Ferragamo:2023}. 

In this work, we aim to move another step further -- studying Deep Learning architectures that could infer the 2D total mass density maps from X-ray, SZ and stellar density  (galaxy members) observations. As a first approach, we limit our application to only studying the idealised cases with simple mock maps (see next section for more details about the maps). Therefore, we ignore the impact of angular resolution, point source contamination, noise and other instrumental effects that characterise real observations.   

\section{Dataset and training}
\subsection{Mock maps}
\label{sec-1}
We use the results of the  hydrodynamical simulations from  {\sc The Three Hundred Project} \cite{Cui:2018}, which correspond to   ``zoom-in'' spherical regions  centred in the 324 most massive halos of the gravity-only {\sc MultiDark} simulation MDPL2 \cite{Klypin:2016} whose cosmology is given by the parameters inferred by the Planck mission \cite{Planck:parameters}. The size of our 324 regions corresponds to spheres  of radius $15h^{-1}\text{Mpc}$, that were  re-simulated with full baryon physics with particle masses  of  dark matter and gas of $M_{\text{particle}}\sim 10^{8}h^{-1}M_{\odot}$. In addition, the simulations  have been run  with two numerical codes that feature different  star formation, supernovae and black hole feedbacks modeling, which are {\sc Gadget-X} \cite{Cui:2018} and {\sc Gizmo-Simba} \cite{Cui:2022}. Note that the 324 regions not only include the central halos, but there is also plenty of additional groups and filament structure.

\begin{figure}[h]
\centering\includegraphics[width=0.6\textwidth]{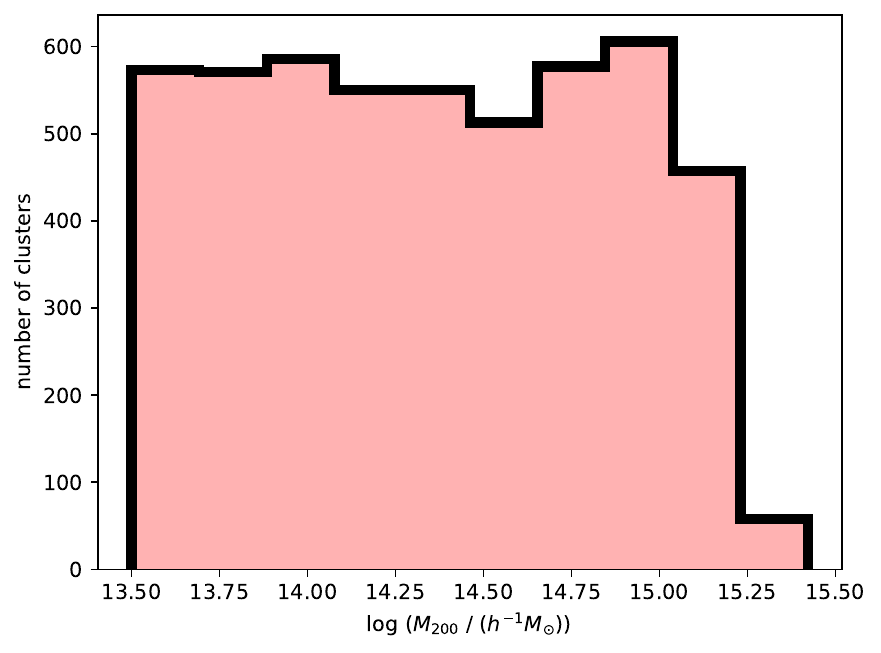}
\caption{Number of galaxy clusters as a function of mass for our selected dataset  at $z\sim0$ from {\sc The Three Hundred}.}
\label{fig-1}       % Give a unique label
\end{figure}

All  halos and sub-halos within these simulations are identified by the {\sc Amiga Halo Finder} (AHF) algorithm \cite{Knollmann:2009} and for this work, distinct halos are selected at redshift $z\sim 0$ with a mass greater than $M_{200}\geq 10^{13.5} h^{-1}M_{\odot}$. $M_{200}$ here means the mass inside a region whose density is 200 times the critical density of the Universe, at the corresponding redshift. Note that the redshift evolution of the scaling laws up to $z\simeq 1$ is negligible and thus, one should not expect a variance in the mass proxies \cite{deAndres:2023}. The mass distribution of the  selected sample is shown in \autoref{fig-1}. We should remark that we have selected our sample of simulated clusters  to follow an almost  uniform mass distribution so that high-mass clusters are not under-represented in the training dataset. Taking this into consideration, our sample is composed of 5,041 different clusters. We have further considered 29 lines of sight (l.o.s) projections and thus, our data set is composed of 146,189 maps per input (tSZ, X-ray, star)  and output view (mass density). 

\begin{figure}[h]
\centering
\includegraphics[width=1\textwidth]{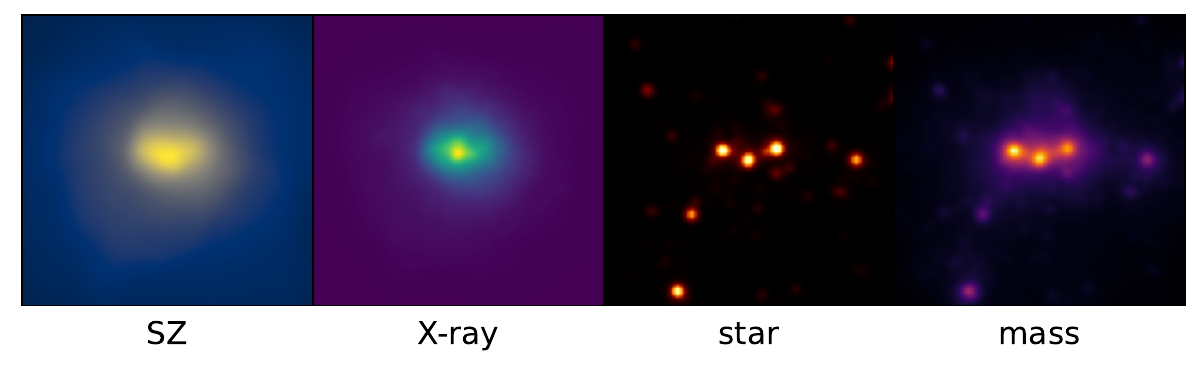}
\caption{Examples of our input mock maps (SZ, X-ray and star) together with the output mass density. The size of all maps is $2\times R_{200}$.}
\label{fig-2}       % Give a unique label
\end{figure}

Regarding our input mock maps, the tSZ maps are created using the publicly available {\sc PYMSZ} package \footnote{https://github.com/weiguangcui/pymsz}, the X-ray maps are the bolometric surface brightness computed  using the AtomDB database \footnote{https://github.com/rennehan/xraylum}, the star maps correspond to the projected star mass density along the observer's l.o.s., and the output maps are the projected total mass density (DM, gas, stars and black holes). Examples of our input and output ground-truth images can be found in \autoref{fig-2}. Furthermore, our maps are reduced to a grid where 80 pixels equals $2\times R_{200}$ for all the maps with a Gaussian smoothing (FWHM) of $\simeq 0.01\times R_{200}$.

\subsection{Model and training}
The model that we have used is based on the U-Net architecture \cite{Ronneberger:2015}, which was developed for the purpose of analysing biomedical images. We have used the particular architecture of \cite{Zbontar:2018} that was also considered in biomedical imaging but we adapted it to the dimensionality of our input dataset. Moreover, that architecture is also modified to account for multiview inference, i.e., it can perform the inference of the mass maps from SZ, X-ray and star data simultaneously. It learns the features that most correlate with mass from each view and optimally combines them for the inference task. The dataset is split into training (90\%) and test (10\%).  Training  1 epoch  of the model  using an A100 NVIDIA GPU takes around 2 minutes. The models were trained for a total of 100 epochs.

\section{Results}
\begin{figure}[h]
\centering
\includegraphics[width=1\textwidth]{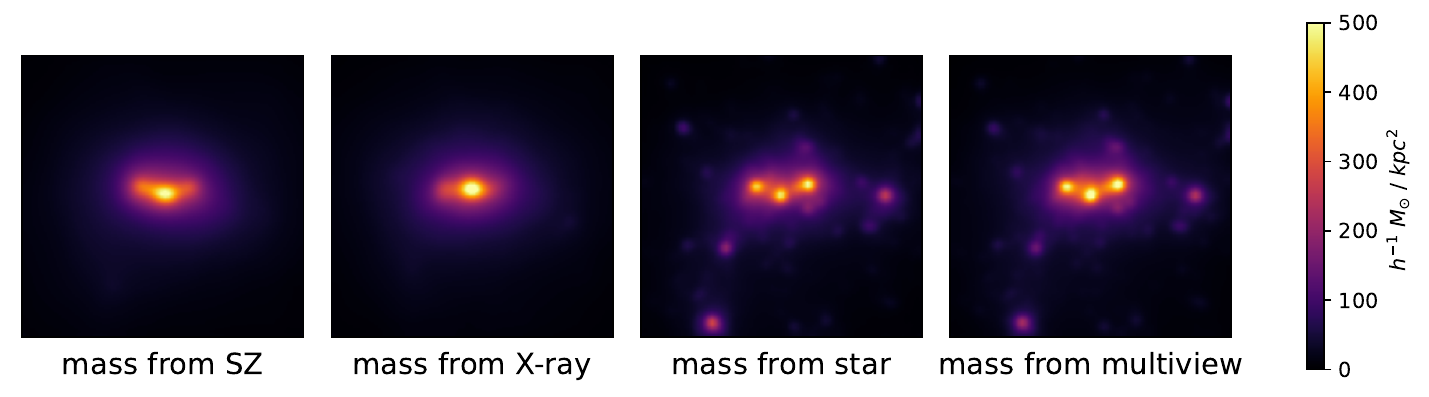}
\caption{Example of our predicted maps varying the inputs of our U-Net model: SZ, Xray, star and multiview.}
\label{fig-3}       % Give a unique label
\end{figure}

 We show examples in \autoref{fig-3} of our predicted maps from 4 different U-Nets: training only with SZ, only with X-ray, only with star and training with all views together in a multiview approach. Note that the ground-truth map is the last map in \autoref{fig-2}. 

\begin{figure}[h]
\centering
\includegraphics[width=1\textwidth]{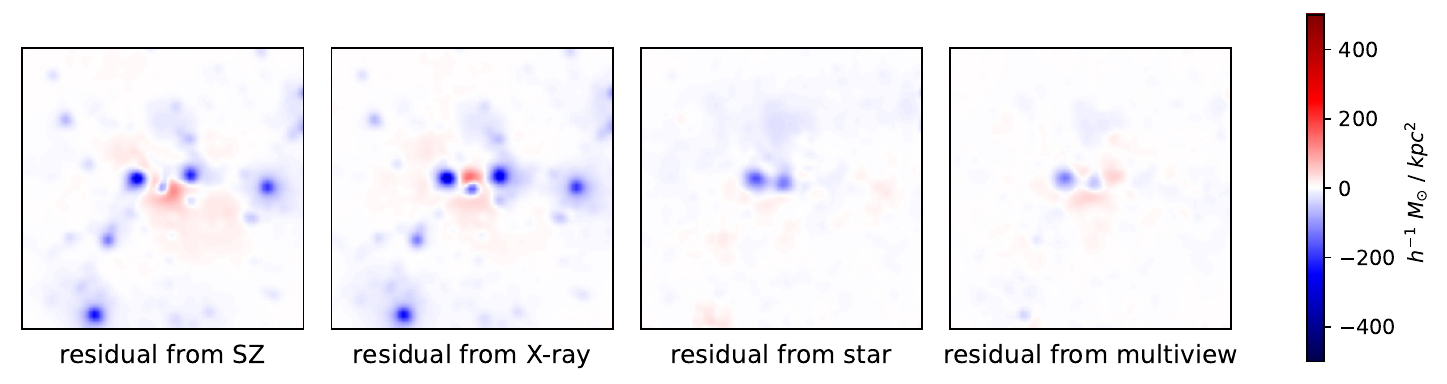}
\caption{Residuals (see \autoref{eq:residual} ) of our predicted maps varying the inputs of our U-Net model: SZ, Xray, star and multiview.}
\label{fig-4}       % Give a unique label
\end{figure}

 As a general result, we can distinguish that the density mass maps from SZ and from X-ray are smoother than those maps predicted from the stars and from multiview. This can be explained by \autoref{fig-2}, where the SZ and X-ray maps do not contain mass distribution information at small scales and therefore, the models that were trained considering only those maps will not predict those small structures. Conversely, star maps contain galaxy members that can track better the positions  of the total mass density peaks. To further study this, we have computed the residuals defined as:
 \begin{equation}\label{eq:residual}
     \text{residuals} = \text{predicted density map}-\text{ground-truth density map .}
 \end{equation}
 An example of the residuals can be seen in \autoref{fig-4}. In this figure, we highlight in blue colour that in the case of SZ and X-ray input maps there are more under-predicted substructures. In our example, for stars, there is a lack of signal in the central region and the residuals are not the substructures that are present in SZ and X-ray residual maps. The advantage of considering the multichannel U-Net model is that the predictions are more accurate due to the fact that the model can point to the substructures from the star maps and it also utilises the information from SZ and X-ray maps to improve the inference of the mass density maps.

\section{Conclusions}
In this work, we have considered the U-Net model to infer mass density maps of galaxy clusters from idealised mock observations of X-ray, SZ and stars. We have shown that the U-Net, which comes from biomedical imaging research, can be generalised for  this purpose from single or multiple input maps. To obtain the location of mass density peaks, one has to consider galaxy member information as depicted in \autoref{fig-3} and \autoref{fig-4}, while the predictions from ICM tracers yield a smooth total mass distribution without substructures. However, to get the most accurately predicted mass density maps, one has to use the multiview approach. This model can be easily generalised to combine as many input maps as possible, for instance, different X-ray bands, stellar luminosity in different photometric bands, etc,  being the only limitation the computational resources required.  

Nevertheless, the limitations of Deep Learning for addressing this problem could come from several sources: a) Firstly, these models need simulated data to be trained, which assumes different cosmological and astrophysical frameworks. This problem can be addressed by training our model with  combined  data from different cosmological simulations accounting for a wide range of plausible physics, as considered by the CAMELS collaboration \cite{CAMELS}; b) Secondly, the models are trained with simulated data without considering instrumental impacts and therefore, they cannot be directly applied to real data. This is a small limitation that can technically be removed by adding the instrumental impact to the clean maps as considered by \cite{deAndres:2022}. However, the models need to be retrained with mock data matching the particular characteristics of each instrument. In subsequent publications, we will address these limitations and will show mathematically and statistically how different  the predictions of our models are from the ground-truth maps, quantifying the power of using Deep Learning for inferring mass density maps directly from observables.

\end{document}